 \documentclass[epj]{svjour}
 \usepackage{graphicx}
 \usepackage{amsmath}
\begin{document}
\title{Phenomena at the QCD phase transition in nonequilibrium chiral fluid dynamics (N$\chi$FD)}
\author{Marlene Nahrgang\inst{1} \and Christoph Herold\inst{2}
}                     
%
%
 \institute{Department of Physics, Duke University, Durham, North Carolina 27708-0305, USA \and School of Physics, Suranaree University of Technology, 111 University Avenue, Nakhon Ratchasima 30000, Thailand}
%
%
\abstract{ Heavy-ion collisions performed in the beam energy range accessible by the NICA collider facility are expected to produce systems of extreme net-baryon densities and can thus reach yet unexplored regions of the QCD phase diagram. Here, one expects the phase transition between the plasma of deconfined quarks and gluons and the hadronic matter to be of first order. A discovery of the first-order phase transition would as well prove the existence of the QCD critical point, a landmark in the phase diagram. In order to understand possible signals of the first-order phase transition in heavy-ion collision experiments it is very important to develop dynamical models of the phase transition. Here, we discuss the opportunities of studying dynamical effects at the QCD first-order phase transition within our model of nonequilibrium chiral fluid dynamics.
} 

 \authorrunning{M.~Nahrgang, C.~Herold}
 \titlerunning{Phenomena at the QCD phase transition in N$\chi$FD}
\maketitle
\section{Introduction}
\label{intro}
Heavy-ion collisions at highest beam energies as performed at the LHC have created the quark-gluon plasma (QGP), which consists of deconfined quarks and gluons, in the laboratory. This liberation of color degrees of freedom at high temperatures is indicated in the change of thermodynamic quantities as calculated from lattice QCD \cite{Borsanyi:2013bia,Bazavov:2014pvz}. At these high energies the net-baryon density in the center region of the collision is very close to zero. Here, lattice QCD calculations have shown that the transition between the QGP and the confined hadronic matter at low temperatures is an analytic crossover  \cite{Aoki:2006we}. At finite baryo-chemical potentials lattice QCD calculations become notoriously difficult due to the fermionic sign problem \cite{Aarts:2014fsa}. 
In order to learn about some aspects of QCD at low temperatures and finite densities one can study low-energy effective models. It is a  generic feature of effective models without vector coupling in mean-field \cite{Scavenius:2000qd,Schaefer:2004en,Ratti:2005jh,Schaefer:2007pw} or beyond \cite{Pawlowski:2005xe,Skokov:2010wb,Skokov:2010sf,Herbst:2010rf,Fischer:2012vc,Fischer:2014ata} to find a phase transition of first-order at high baryonic densities. As a consequence the line of first-order phase transition ends in a critical point, a singular point of a second-order phase transition. If the vector coupling is increased, however, the critical point moves to lower temperatures and higher densities until it disappears completely \cite{Kitazawa:2002bc,Sasaki:2006ws,Fukushima:2008wg,Bratovic:2012qs}. It would be an interesting task to constrain the value of the vector coupling by comparison to either experimental data or lattice QCD calculations \cite{Kunihiro:1991qu,Contrera:2012wj,Steinheimer:2014kka}.

The QCD critical point is of special interest, because fluctuations of conserved quantities and order parameter fields diverge at second-order phase transitions according to their universal behavior. If the system created in a heavy-ion collision can be described thermodynamically in the vicinity of the critical point, large event-by-event fluctuations should be seen in measured multiplicities, such as net-charge or net-proton multiplicities \cite{Halasz:1998qr,Stephanov:1998dy,Stephanov:1999zu,Hatta:2002sj,Stephanov:2008qz}. It is well known, however, that dynamical systems experience the phenomenon of critical slowing down \cite{Berdnikov:1999ph}, which arises due to the divergence of the relaxation time near a critical point. It is currently debated how the fast dynamics of the expansion of the QGP will affect critical signatures in fluctuation observables \cite{nahrgangcpod2014,Mukherjee:2015swa}.

Another possibility of proving the existence of the critical point is to look for signals of the first-order phase transition at large baryonic densities, as will be reached by upcoming heavy-ion experiment facilities like NICA in Dubna and FAIR at GSI. At a first-order phase transition two thermodynamic phases coexist and are separated by a potential barrier associated with the latent heat. In thermodynamic systems the equation of state for a first-order phase transition is obtained by a Maxwell construction for the pressure of the coexistence region which connects the pressures of the high and the low temperature phases. By this construction the system is thermodynamically stable and the speed of sound vanishes. The equation of state is very soft and various observables have been proposed to be sensitive to it \cite{Kolb:2000sd,Kolb:2003dz}. If a system,  however, goes dynamically through a first-order phase transition, it might be trapped in the metastable state, a phenomenon called supercooling. The metastable state can decay via nucleation or for large nucleation times via spinodal decomposition \cite{Csernai:1995zn,Zabrodin:1998dk,Keranen:2002sw,Nahrgang:2011vn}. Large inhomogeneities, domain formation and an enhancement of the low-momentum modes are expected in these systems \cite{Mishustin:1998eq,Randrup:2009gp,Randrup:2010ax,Steinheimer:2012gc}. 

In order to understand the impact of the dynamics on the phase transition phenomena it is of crucial importance to develop models which couple the phase transition to the expansion of a heavy-ion collision. 
In the present work, we will discuss the capabilities of nonequilibrium chiral fluid dynamics to address these questions. We focus on the first-order phase transition, which lies in the region of the phase diagram expected to be covered by the NICA facility.

\section{Nonequilibrium chiral fluid dynamics (N$\chi$FD)}\label{sec:1}

In the last couple of years the model of N$\chi$FD for the explicit propagation of fluctuations coupled to a dynamic expansion of a heavy-ion collision was developed \cite{Nahrgang:2011mg} showing that nonequilibrium effects can have an important influence on the evolution of the fluctuations of the order parameter of chiral symmetry \cite{Nahrgang:2011vn,Nahrgang:2011mv,Herold:2013bi,Herold:2013qda,Herold:2014zoa}. This approach starts from a low-energy effective model, such as the quark-meson (QM) or Polyakov-loop extended quark-meson (PQM) model \cite{Scavenius:2000qd,Schaefer:2004en,Ratti:2005jh,Schaefer:2007pw}. The equation of motion for the sigma field are derived within the framework of the two-particle irreducible effective action as
\begin{equation}
 \partial_{\mu}\partial^{\mu}\sigma+\eta_{\sigma}(T)\partial_t \sigma+\frac{\partial (U+\Omega_{\bar qq})}{\partial\sigma}=\xi_{\sigma}\, .
\label{eq:eoms1}
\end{equation}
Here, $U$ is the classical chiral potential including a term for explicit chiral symmetry breaking and 
\begin{align} 
\Omega_{\rm q\bar q}&=-2 N_f N_c T\int\frac{\mathrm d^3 p}{(2\pi)^3} \left\{\ln\left[1+\mathrm e^{-\frac{E_{\rm q}-\mu}{T}}\right]\right.\\\nonumber 
                    &\phantom{-2 N_f N_c T\int\frac{\mathrm d^3 p}{(2\pi)^3}}+\left.\ln\left[1+\mathrm e^{-\frac{E_{\rm q}+\mu}{T}}\right]\right\}\, .
\end{align}
is the thermodynamic potential for the fermionic degrees of freedom in mean-field approximation. The energy of the constituent quarks is given by $E_{\rm q}=\sqrt{p^2+g^2\langle \sigma\rangle^2}$. The damping coefficient $\eta_{\sigma}(T)$ is calculated via the scattering processes of the sigma field with the quarks and antiquarks. 
Assuming that the relaxation time of the quarks and antiquarks is much shorter than that for the order parameter, they constitute a heat bath. By effectively integrating out these degrees of freedom the stochastic noise term $\xi_{\sigma}$ is generated from the information loss of this sector. It is approximated as Gaussian white noise and the magnitude of the variance depends via the dissipation-fluctuation theorem on the damping coefficient. It is crucial for this approach to note that the dynamics of the fluctuations of the order parameter is not deterministic.

Since the Polyakov-loop $\ell$ is an effective field, the dynamics cannot be calculated directly and we use a phenomenologically motivated stochastic relaxation equation according to
\begin{equation}
  \eta_{\ell}\partial_t \ell T^2+\frac{\partial ({\cal U}+\Omega_{\bar qq})}{\partial\ell}=\xi_{\ell}\, .
\label{eq:eoms2}
\end{equation}
For the same reasons the damping term $\eta_{\ell}$ cannot be derived from the underlying model either. We use a value of $\eta_{\ell}=5/$fm, and the results turn out to be rather independent of the precise value. ${\cal U}$ is the polynomial form of the temperature-dependent Polyakov-loop potential.

The equations of motion of the quark-antiquarks are coarse-grained to yield the fluid dynamical equations of energy-momentum conservation with the pressure $ p = -\Omega_{q\bar q}(\sigma; T,\mu)$.
Since, however, the fluid sector is coupled to the propagation of the sigma field there will be an exchange of energy and momentum in a finite system considered here. Therefore, the fluid dynamical equations are augmented with a source term
\begin{align}
\label{eq:fluidT}
\partial_\mu T^{\mu\nu}&=S^\nu\, ,\\
\partial_\mu N_{\rm q}^{\mu}&=0\, .
\label{eq:fluidN}
\end{align}
Due to the stochastic evolution of the sigma field the source term
\begin{equation}
S^\nu=-\partial_\mu\ T_\sigma^{\mu\nu}\, .
\end{equation}
is also of stochastic nature and fluctuations in the sigma field couple to the conserved densities. 

  \begin{figure}
  \centering
   \includegraphics[width=0.48\textwidth]{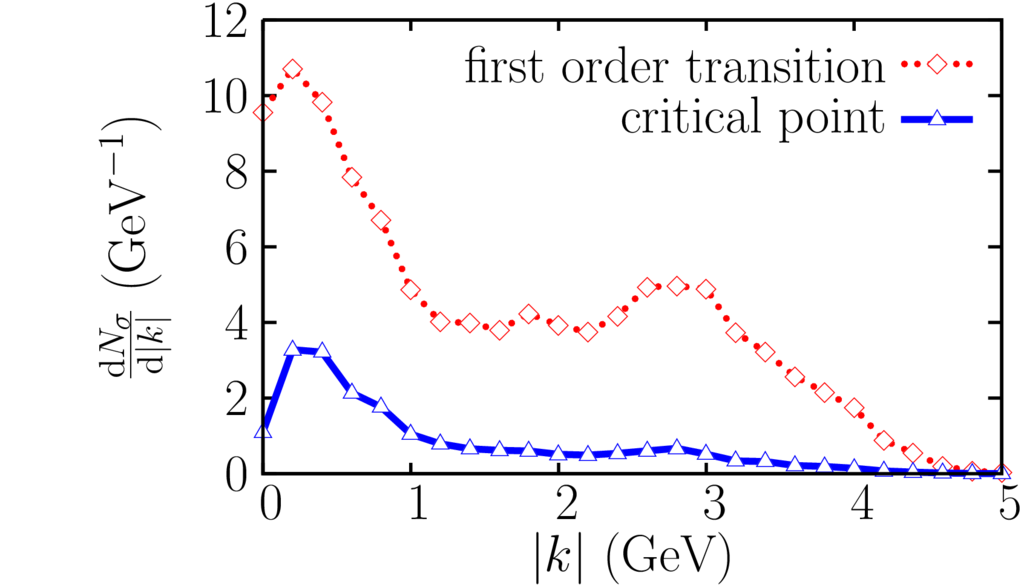}\\
  \vspace{0.2cm}
 
   \includegraphics[width=0.48\textwidth]{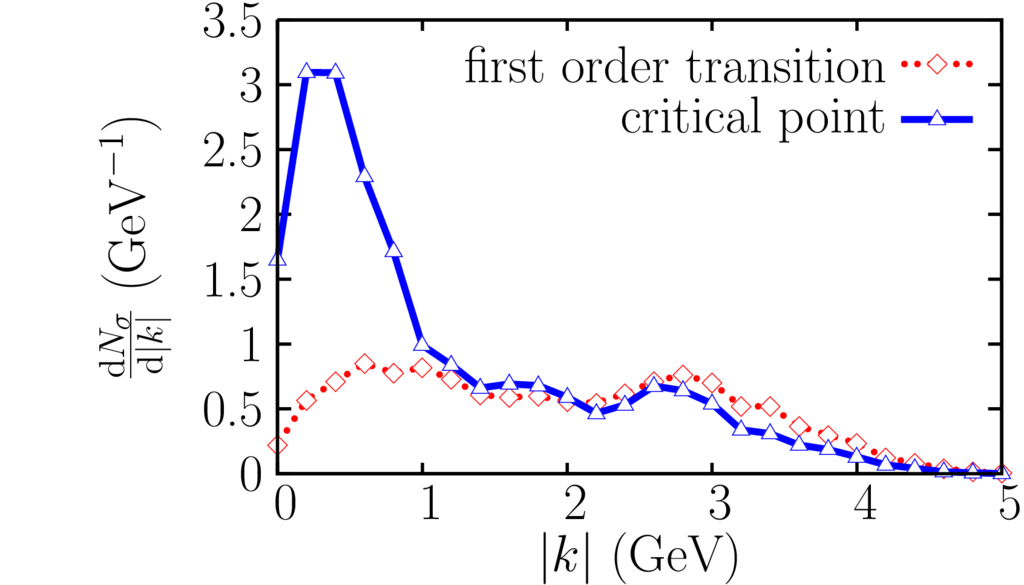}
   \caption{The soft modes of the sigma field for times of $t=12$ fm/c (upper panel) and for $t=24$ fm/c after equilibration of the system in a box calculation \cite{Herold:2013bi}.}
   \label{fig:time}
  \end{figure}
 
   \begin{figure*}
   \centering
   \resizebox{0.23\textwidth}{!}{\includegraphics{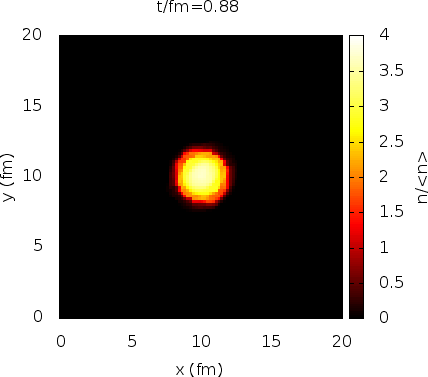}}
   \resizebox{0.23\textwidth}{!}{\includegraphics{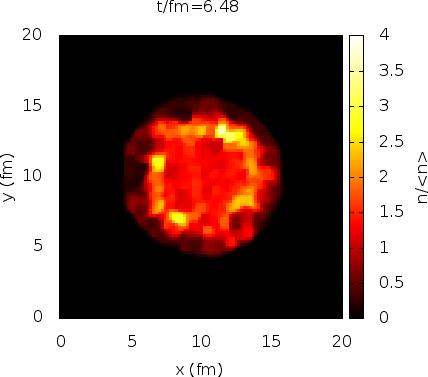}}
  \resizebox{0.23\textwidth}{!}{\includegraphics{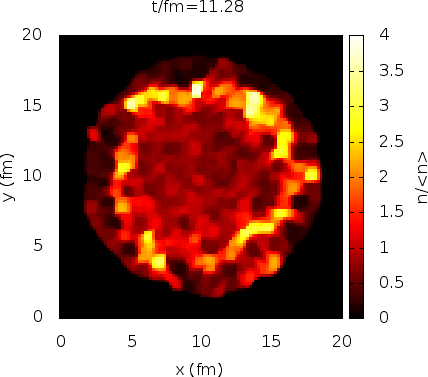}}
  \resizebox{0.23\textwidth}{!}{\includegraphics{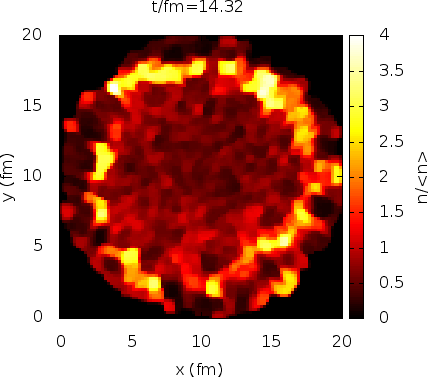}}\\
  \vspace{0.2cm}
  
  \resizebox{0.23\textwidth}{!}{\includegraphics{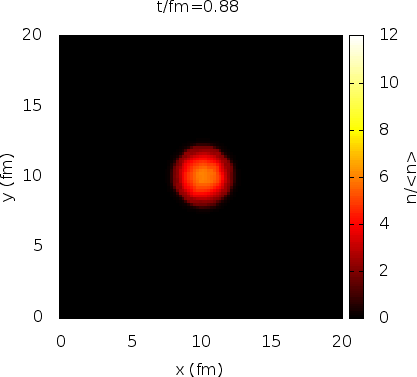}}
  \resizebox{0.23\textwidth}{!}{\includegraphics{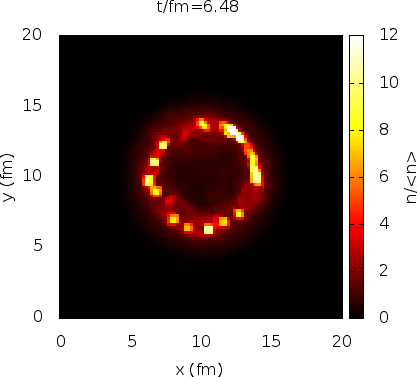}}
  \resizebox{0.23\textwidth}{!}{\includegraphics{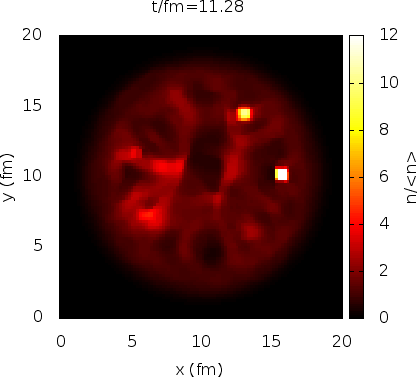}}
  \resizebox{0.23\textwidth}{!}{\includegraphics{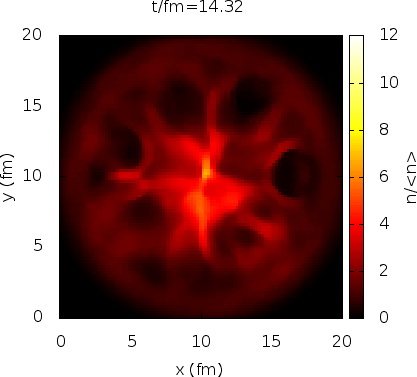}}
  \caption{Snapshot of the evolution of the domain formation and decay through a first-order phase transition at different times. Shown is the normalized local net-baryon density in the $x-y$-plane at $z=0$. The top row shows the dynamics of a PQM equation of state and the lower row that of the QH equation of state.}
  \label{fig:domform}
  \end{figure*}

We test our model first in a static box set-up. The sigma field is initialized at its equilibrium value for a temperature well above the respective transition temperature via the solution of the gap equation in mean-field approximation
\begin{equation}
\label{eq:equilibrium}
\left. \frac{\partial(U+\Omega_{\bar qq}) }{\partial\sigma}\right |_{\sigma=\sigma_{\rm eq}}=0\, .
\end{equation}
The matter in the static box is itself initialized at a temperature below the transition temperature. At the beginning of the evolution, there is thus a large nonequilibrium situation. In this scenario, also known as a sudden temperature quench, we can study the relaxational evolution of the coupled system.
As the system crosses the phase transition dynamical fluctuations are generated. We plot a  measure for the intensity of the sigma field fluctuations $N_\sigma$ as a distribution for the different modes in Fig.~\ref{fig:time}. Two scenarios are shown, including the critical point and the first-order phase transition. The upper plot shows the results after $t=3$~fm/c for the critical point scenario and after $t=12$~fm/c for the scenario with a first-order phase transition. The lower plot shows both scenarios after $t=24$~fm/c. First, one observes that the relaxation of the scenario including the first-order phase transition is significantly delayed compared to the critical point scenario and the fluctuation intensity of the low modes is enhanced only during the relaxation process. Second, the critical point scenario shows an enhancement in the equilibrated state which is similar to the enhancement during the expansion. Although the relaxation time near the critical point is larger than in the crossover regime the barrier in a first-order scenario affects the dynamics of the system more strongly when comparing equilibrium versus nonequilibrium.

We like to point out that the current framework is very flexible in incorporating different types of low-energy effective models. We recently investigated a chiral model with a dilaton field \cite{Sasaki:2011sd,Herold:2014zoa}.

\section{Domain formation}

In N$\chi$FD fluctuations are generated dynamically which means that even for smooth initial conditions spinodal instabilities may arise during the evolution through the phase transition, as was shown in \cite{Herold:2013qda}. The results in this section are obtained in an expanding system with spherical initial conditions of a smooth energy density and net-baryon density profile exponentially decreasing at the edges. The order parameter fields are initially in equilibrium at the local temperature and baryochemical potential.

\begin{figure}[t]
  \centering
   \includegraphics[width=0.38\textwidth,angle=-90]{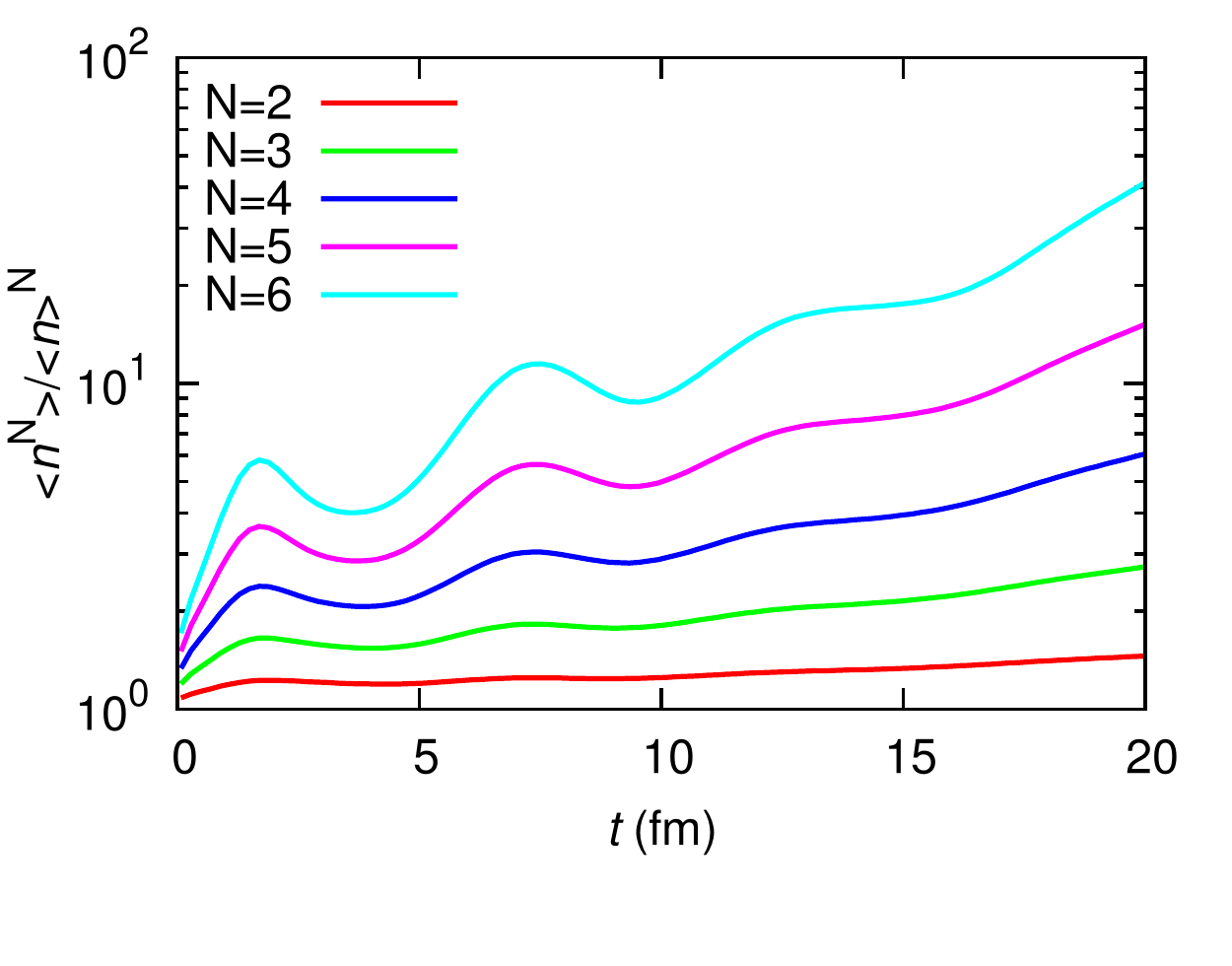}\\
   \includegraphics[width=0.38\textwidth,angle=-90]{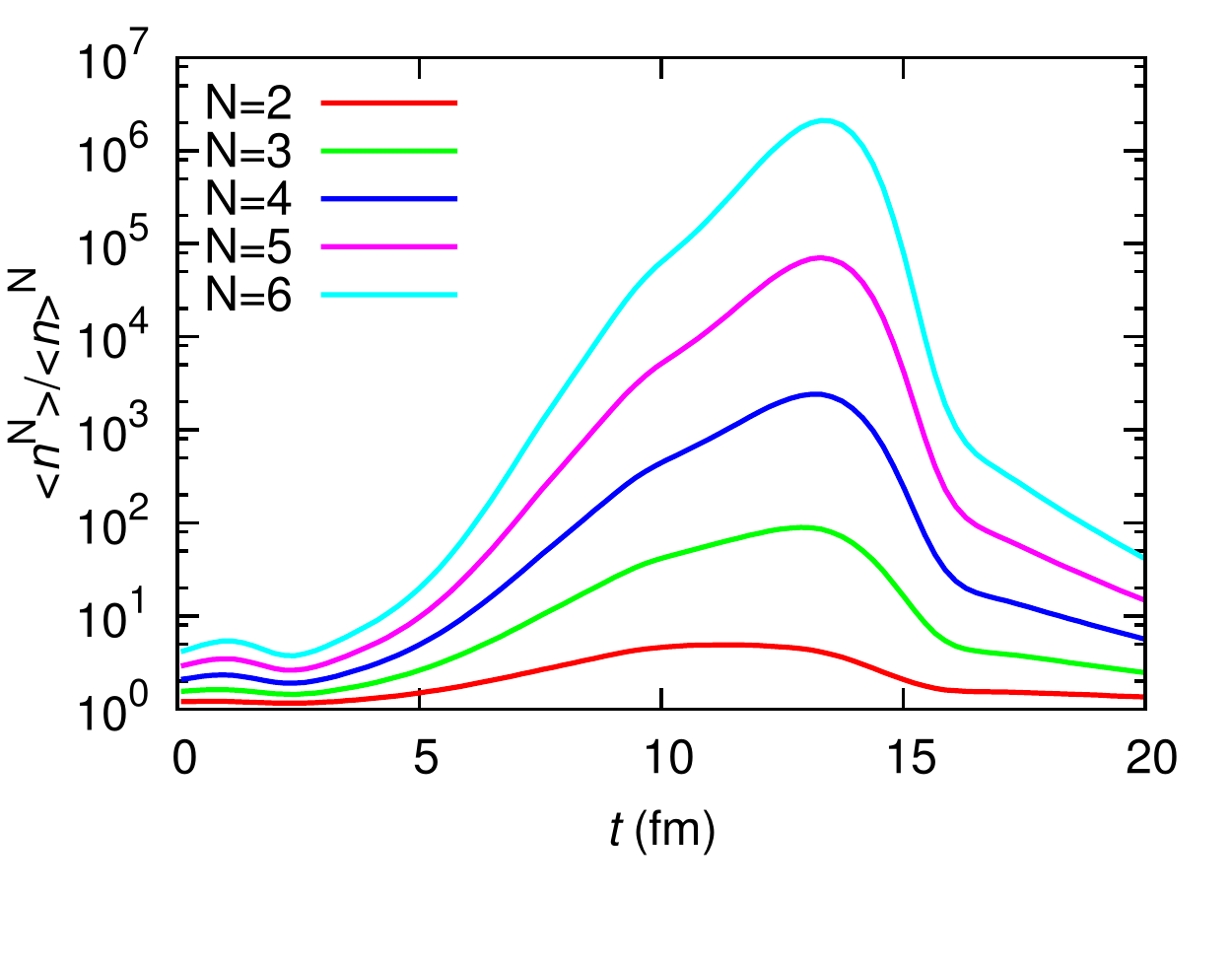}
  \label{fig:moments}
  \caption{Time evolution of the moments of the net-baryon number density distribution over space normalized by the respective power of the mean. The top panel shows the results for the PQM model, the bottom panel shows results for the QH model.}
  \end{figure}

In Fig.~\ref{fig:domform} we compare snapshots of the evolution of the net-baryon number density at different instances of time ($t\, c/\text{fm} =0.88, 6.48, 11.28, 14.32$ from left to right) in the $x-y$-plane at $z=0$. We normalize the local net-baryon number density by the spatial average of the same quantity. In the upper row the effective model is the PQM model, while the lower row shows an evolution according to an effective chiral quark-hadron (QH) model. This model is similar to the SU($3$) non-linear sigma model with quark degrees of freedom discussed in \cite{Dexheimer:2009hi}. 
The hadronic fields include the baryon octet, the vector-isoscalar $\omega$ and $\phi$ and the scalar-isoscalar $\sigma$ and $\zeta$ mesonic condensates. We do not include the $\rho$- and the $\delta$-meson, which are important for isospin asymmetric matter as in neutron stars. The model in  \cite{Dexheimer:2009hi} includes a phenomenological field $\Phi$, which suppresses the baryonic degrees of freedom for high temperatures and densities and the quark degrees of freedom at low temperatures and densities.
The equation of state which is obtained within the (P)QM model is that of a liquid-gas phase transition: the pseudocritical pressure increases with the temperature $\partial p_c/\partial T>0$ and vanishes at zero temperature. Consequently, dense quark matter can coexist with the vacuum at zero temperature \cite{Steinheimer:2013xxa}. We know, however, that dense quark matter at low temperatures coexists with compressed nuclear matter. At high baryonic densities the partonic phase during a heavy-ion collision is expected to be shorter than at zero net-baryon density \cite{Cassing:2009vt,Petersen:2008dd}, and the impact of the hadronic interactions on the global observables is more significant. It is, therefore, extremely important to properly treat the hadronic degrees of freedom at low temperatures within N$\chi$FD for a fully realistic dynamical model of heavy-ion collisions near the phase transition.
The QH model improves this situation since the hadronic degrees of freedom provide a nonzero pressure at low temperatures and the pseudocritical pressure decreases with the temperature $\partial p_c/\partial T<0$ \cite{Hempel:2013tfa}. The QH phase transition is thus phenomenologically very different from the liquid-gas phase transition. This equation of state presented in \cite{Dexheimer:2009hi} and used in its equilibrium form for hybrid model simulations of heavy-ion collisions in \cite{Steinheimer:2009nn} reproduces well the phenomenology of saturated nuclear matter, which the PQM model does not. We can directly observe the differences between these two models by comparing the upper and the lower row of Fig.~\ref{fig:domform}. Starting from smooth spherical initial conditions the system dynamically develops spatial inhomogeneities in the coexistence region. This is most clearly observed in the second plot from the left, upper and lower row, where regions of particularly high densities with respect to the average, these are droplets of quark degrees of freedom, and regions of low densities, are found. This pattern develops in both equations of state. The most prominent difference between the PQM and the QH equation of state lies in the late stage evolution. Due to the vanishing pressure at low temperatures quark droplets remain stable in the case of the PQM equation of state as can be seen in the two right plots of the upper row in Fig.~\ref{fig:domform}. For the QH equation of state the quark droplets decay in the final stages of the evolution. This behavior can be quantified by looking at the time evolution of the normalized moments of the spatial net-baryon density distribution
\begin{align}
 \langle n^N\rangle&=\int{\rm d}^3 x n(x)^N P_n(x)\\ {\rm with}\quad P_n(x)&=\frac{n(x)}{\int{\rm d}^3 x n(x)}\nonumber \, .
 \end{align}
 This is shown in Fig.~\ref{fig:moments} for the PQM equation of state in the top panel and for the QH equation of state in the bottom panel. One sees the continuous increase for the PQM equation of state (with some oscillations) and the dynamical formation and decay of the spatial inhomogeneities for the QH equation of state. 
 
\section{Conclusions}

The model of nonequilibrium chiral fluid dynamics offers the unique possibility to study the dynamics of fluctuations at the phase transition embedded in a realistic expansion of the matter created in a heavy-ion collision. Especially at a first-order phase transition it is important to properly take into account the nonequilibrium effects which arise due to the fast dynamics. While expanding a background fluid of quarks and antiquarks we propagated the order parameter fields, like the sigma field or the Polyakov loop, explicitly. These two sectors are coupled via the damping coefficient and the noise term, the equation of state and the stochastic source in the fluid dynamical equations ensuring energy and momentum conservation of the total system.

This model is able to reproduce the qualitative features of the nonequilibrium effects at the first-order phase transition such as the supercooling and subsequent decay of the metastable state leading to domain formation at the phase transition.

In order to make quantitative predictions for possible signals of the first-order phase transition, like for example enhanced flow coefficients due to the spatial irregularities of domain formation or enhanced deuteron over proton yields, the current model needs to be improved such that the hadronic phase is treated correctly. Finally, of course realistic initial conditions including the initial baryon stopping and the late-stage hadronic interactions need to be included to make reliable predictions for the upcoming NICA facility.

\section{Acknowledgments}
M.N. acknowledges support from a fellowship within the Postdoc-Program of the German Academic Exchange Service (DAAD). This work was supported by the U.S. department of Energy under grant DE-FG02-05ER41367. C.H. acknowledges funding by Suranaree University of Technology (SUT) and CHE-NRU (NV.10/2558) project. The authors are grateful to V.~Dexheimer, J.~Steinheimer, J.~Randrup and V.~Koch for fruitful discussions.

%

\end{document}